\RequirePackage{fix-cm}
\documentclass[smallextended]{svjour3}       
\smartqed  
\usepackage{graphicx}
\usepackage{epstopdf}
\usepackage{amssymb}
\usepackage{amsmath}
\usepackage{graphicx}
\usepackage{subfigure}
\usepackage{algorithm}
\usepackage{url}
\DeclareMathOperator*{\argmin}{arg\,min}

\begin{document}

\title{A Survey of Point-of-interest Recommendation in Location-based Social Networks
}

\titlerunning{Survey for POI Recommendation }        

\author{Shenglin Zhao, Irwin King, and Michael R.~Lyu
}

\institute{Shenglin Zhao, Irwin King, and Michael R.~Lyu \at 
Department of Computer Science \& Engineering\\ The Chinese University of Hong Kong,
Shatin, N.T., Hong Kong\\
\email{\{slzhao, king, lyu\}@cse.cuhk.edu.hk}
}

\date{Received: date / Accepted: date}

\maketitle

\begin{abstract}
Point-of-interest (POI) recommendation that suggests new places for users to visit arises with the popularity of location-based social networks (LBSNs). Due to the importance of POI recommendation in LBSNs, it has attracted much academic and industrial interest. In this paper, we offer a systematic review of this field, summarizing the contributions of individual efforts and exploring their relations. We discuss the new properties and challenges in POI recommendation, compared with traditional recommendation problems, e.g., movie recommendation. Then, we present a comprehensive review in three aspects: influential factors for POI recommendation, methodologies employed for POI recommendation, and different tasks in POI recommendation. Specifically, we propose three taxonomies to classify  POI recommendation systems. First, we categorize the systems by the influential factors check-in characteristics, including the geographical information, social relationship, temporal influence, and content indications. Second,  we categorize the systems by the methodology, including systems modeled by fused methods and joint methods. Third, we categorize the systems as general POI recommendation and successive POI recommendation  by subtle differences in the recommendation task whether to be bias to the recent check-in. For each category, we summarize the contributions and system features, and highlight the representative work. Moreover, we discuss the available data sets and the popular metrics. Finally, we point out the possible future directions in this area and conclude this survey. 
\end{abstract}

\section{Introduction}
\begin{figure}[h!]
\center
\includegraphics[width=3in, height=2in]{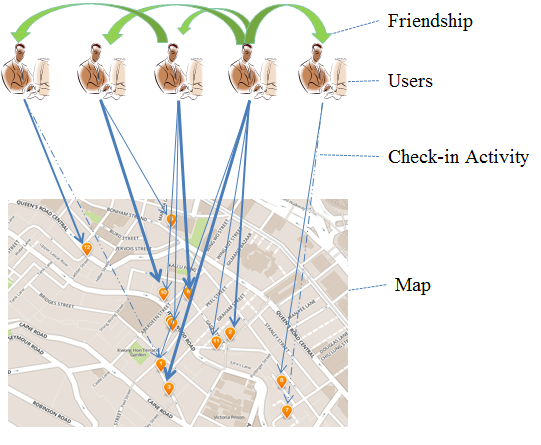}
\caption{A typical LBSN (The line weight demonstrates check-in frequency; the more weighted the line is, the more frequently one user visits the POI. Dash line is used to show the one-time check-in. As shown, users visit POIs differently, showing specific preferences.)}
\label{fig:lbsn}
\end{figure}

Location-based social networks (LBSNs) such as Foursqaure, Facebook Places, and Yelp are popular now owning to the explosive increase of smart phones. Sharp increase of smart phones arouses prosperous online LBSNs. Until June 2016, Foursquare has collected more than 8 billion check-ins and more than 65 million place shapes mapping businesses around the world; over 55 million people in the world use the service from Foursquare each month\footnote{https://foursquare.com/about}. LBSNs collect users' check-in information including visited locations' geographical information (latitude and longitude) and users' tips at the location. LBSNs also allow users to make friends and share information.  Figure~\ref{fig:lbsn} demonstrates a typical LBSN, exhibiting the interactions (e.g., check-in activity) between users and POIs, and interactions (friendship) among users. In order to improve user experience in LBSNs,  point-of-interest (POI) recommendation is proposed that suggests new places for users to visit from mining users' check-in records and social relationships. 

POI recommendation is  one of the most important tasks in LBSNs, which helps users discover new interesting locations in the LBSNs.  POI recommendation typically mines users' check-in records, venue information such as categories, and users' social relationships to recommend a list of POIs where users most likely check-in in the future. POI recommendation not only improves user viscosity to LBSN service providers, but also benefits advertising agencies with an effective way of launching advertisements to the potential consumers.  Specifically, users can explore nearby restaurants and downtown shopping malls in Foursquare. Meanwhile, the merchants are able to make the users to easily find them through POI recommendation. Owning to the convenience to users and business opportunities for merchants, POI recommendation attracts intensive attention and a bunch of POI recommendation systems have been proposed recently~\cite{lian2015content,sang2015activity,wang2015semantic,wang16spore,yin2016joint,zhang2015orec,zhao2016stellar}.

POI recommendation is a branch of recommendation systems, which indicates to borrow ideas for this task from conventional recommendation systems, e.g., movie recommendation. We suffice to make use of conventional recommendation system techniques, e.g., collaborative filtering methods. However, the specific fact that location concatenates the physical world and the online networking services, arouses new challenges  to the traditional recommendation system techniques. We summarize some confronting  challenges  as follows,
\begin{enumerate}
\item Physical constraints: Check-in activity is limited by physical constraints, compared with shopping online from Amazon and watching movie in Netflix. For one thing, users in LBSNs check-in at geographically constrained areas; for another, shops regularly provide services in some limited time. Such physical constraints make the check-in activity in LBSN exhibit significantly spatial and temporal properties~\cite{bhargava2015and,cheng2011exploring,gao2012gscorr,Gao2015Addressing,rhee2011levy,yang2015modeling,yin16discovering}. 
\item Complex relations: For online social media services such as Twitter and Facebook, location is a new object, which yields new relation between locations~\cite{yuan2012discovering}, between users and locations~\cite{gao2012exploring,sattari2012geo,yin2015dynamic}. In addition, location sharing activities alter relations between users since people are apt to make new friends with geographical neighbors~\cite{Scellato2010Distance,Scellato2011Exploiting}. 
\item Heterogeneous information: LNSNs consist of different kinds of information, including not only check-in records, the geographical information of locations, and venue descriptions but also users' social relation information and media information (e.g., user comments and tweets). The heterogeneous information depicts the user activity from a variety of perspectives~\cite{wang2015predicting,Wang2015Regularity,Zhang2014Transferring}, inspiring POI recommendation systems of different kinds~\cite{liu2015general,liu2014exploiting,liu2013point,noulas2012mining,sang2015activity,wang2015semantic,yuan2014graph}.
\end{enumerate}

\begin{table}[t!]
\centering
\caption{Statistics on the literature}
\begin{tabular}
{p{2.3cm} p{1cm} p{1cm} p{1cm} p{1cm} p{1cm}p{1cm} p{1cm}} \hline
Name &     {2010}&{2011}&{2012}&{2013}&{2014}&{2015}&{2016}\\ \hline
\textbf{Conference} \\
{AAAI}        &    &    & 1  &    &    & 1 & 3 \\ 
{IJCAI}       &    &    &    & 1  &    & 1 & 1 \\ 
{ICDE}        &    &    &    &    &    &   &  2  \\ 
{ICDM}        &    &    & 1  &    &  1 & 2 &    \\ 
{WWW}         &    &    &    &    &    & 1 &    \\ 
{KDD}         &    &  1  &    &  2 &  1 & 1 & 2  \\
{SIGIR}       &    & 1  &    & 1  &    & 4 &  \\ 
{SIGSPATIAL}  &  1  &   & 1  &  2 & 1  &   & \\ 
{CIKM}        &    &    &  1 &  1 &  2 & 3 &  \\ 
{RecSys}      &    &    &    &  2 &    & 1 &  \\ 
{SDM}         &    &    &    & 1  &    &   &  \\ 
{Ubicomp}      &    &    &  1 &    &    &   &  \\ 
{ICWSM}       &    &    &  1 &    &    &   &  \\ 
{WSDM}        &    &    &    &  1  &    &   &  \\ 
 &    &    &    &    &    &    &  \\
\textbf{Journal} \\
{TKDE}        &    &    &    &    &    & 1 &  1 \\ 
{TIST}        &    &    &    &    &    & 1 &  1 \\ 
{TOIS}        &    &    &    &    &  1 &   &   \\ 
{TKDD}        &    &    &    &    &    & 1 &   \\
{TSC}         &    &    &    &    &    & 1 &  \\  
{TMM}         &    &    &    &    &    & 2 & 1 \\ 
{DMKD}        &    &    &    &    &    & 1 &  \\ 
{Neurocomputing} &    &    &    &    &    &   & 1 \\
 &    &    &    &    &    &    &  \\
\textbf{Total}& 1  &  2 & 6  & 11 & 6  & 20 & 12 \\
\hline\end{tabular}
\label{tbl:stat}
\end{table}

A bunch of researches are carried out to address this significant but challenging problem---POI recommendation. Ye et al.~\cite{ye2010location} first propose POI recommendation for LBSNs such as  Foursquare and Gowalla. After that, more than 50 papers about the problem are published in top conferences and journals, including SIGKDD, SIGIR, IJCAI, AAAI, WWW, CIKM, ICDM, RecSys, TIST, TKDE, TIST, and so on so forth. Table~\ref{tbl:stat} shows the statistics on the literature. Some similar researches with POI recommendation, such as restaurant recommendation system~\cite{horozov2006using} or location recommendation from GPS trajectories~\cite{zheng2009mining,zheng2011learning,sattari2012geo}, base on the other types of data, beyond our scope. In this survey, we focus on the POI recommendation for LBSNs.  We surpass the latest survey~\cite{yu2015survey} in this field in depth and scope: 1) Yu et al.~\cite{yu2015survey} only categorize the POI recommendation according to the influential factors, while, we show the taxonomies from three perspectives. 2) We incorporate more researches, especially systems established on joint models and some recently published papers. 3) We show the trends and new directions in this field. 
 
\begin{figure}[h!]
\center
\includegraphics[width=4.5in, height=2in]{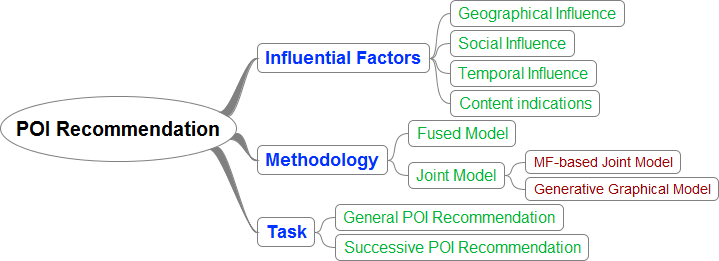}
\caption{Demonstration of taxonomies for POI recommendation}
\label{fig:frame}
\end{figure}

We follow the scheme shown in Fig.~\ref{fig:frame} to reveal academic progress in the area of POI recommendation. We categorize the POI recommendation systems in three aspects: influential factors, methodology, and task. More specifically, we discuss four types of influential factors: geographical influence, social influence, temporal influence, and content indications. In addition, we categorize the methodologies for POI recommendation as fused models and joint models. Moreover, we categorize POI recommendation systems as general POI recommendation and successive POI recommendation according to the subtle difference in task whether to be inclined to the recent check-in.
To report these contents, we organize the remain of this paper as follows. Section~\ref{sec:pd}  reports the problem definition. Section~\ref{sec:inf} demonstrates the influential factors for POI recommendation. Next, Section~\ref{sec:tm} and \ref{sec:tt} show the POI recommendation systems categorized by methodology and task, respectively. 
Then, Section~\ref{sec:pe} introduces data sources and metrics for system performance evaluation. Further,  Section~\ref{sec:tnd}  points out the trends and new directions in the POI recommendation area. Finally,  Section~\ref{sec:con} draws the conclusion of this paper. 

\begin{figure}[h!]
\center
\includegraphics[width=2.5in, height=2.8in]{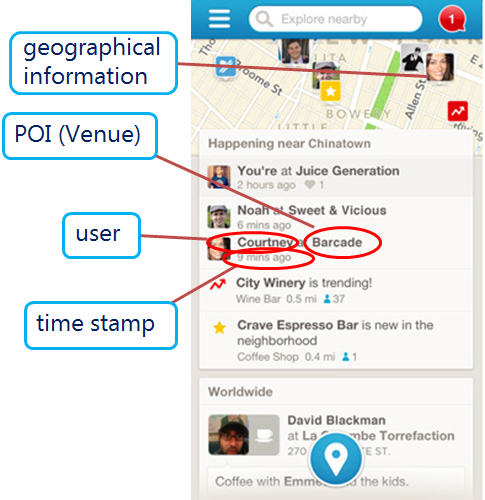}
\caption{Demonstration of check-in information in Foursquare}
\label{fig:checkin}
\end{figure}
\section{Problem Definition}
\label{sec:pd}
POI recommendation aims to mine users' check-in records and recommend POIs for users in LBSNs. Take Foursquare as an example, Figure~\ref{fig:checkin} demonstrates how the check-in information is recorded, including user name, POI, check-in time stamp, and geographical information in the map.
Formally, we define two important terms, i.e., check-in and check-in sequence, as follows. 

\begin{definition}[Check-in]
A check-in is denoted as a triple $\langle u,l,t \rangle$ that depicts a user $u$ visiting POI $l$ at time $t$. 
\end{definition}

\begin{definition}[Check-in sequence]
 A check-in sequence is a set of check-ins of user $u$, denoted as $S_u = \{ \langle l_1, t_1 \rangle, \dots , \langle l_n, t_n \rangle\} $, where $t_i$ is the check-in time stamp.  For simplicity, we denote $S_u=\{l_1, \dots, l_n \}.$
\end{definition}

POI recommendation aims to recommend a user a list of unvisited POIs via mining the check-in records. Hence the problem of POI recommendation can be defined as follows.
\begin{definition}[POI recommendation] Given all users' check-in sequences $S$, POI recommendation aims to recommend a POI list $S_N$ for to each user $u$. Here $S$ is a collected check-in sequence set, contain all sequences $S_u$ for all users.  

\end{definition}

\section{Taxonomy by Influential Factors}
\label{sec:inf}
\begin{figure}[h!]
\center
\includegraphics[scale=0.45]{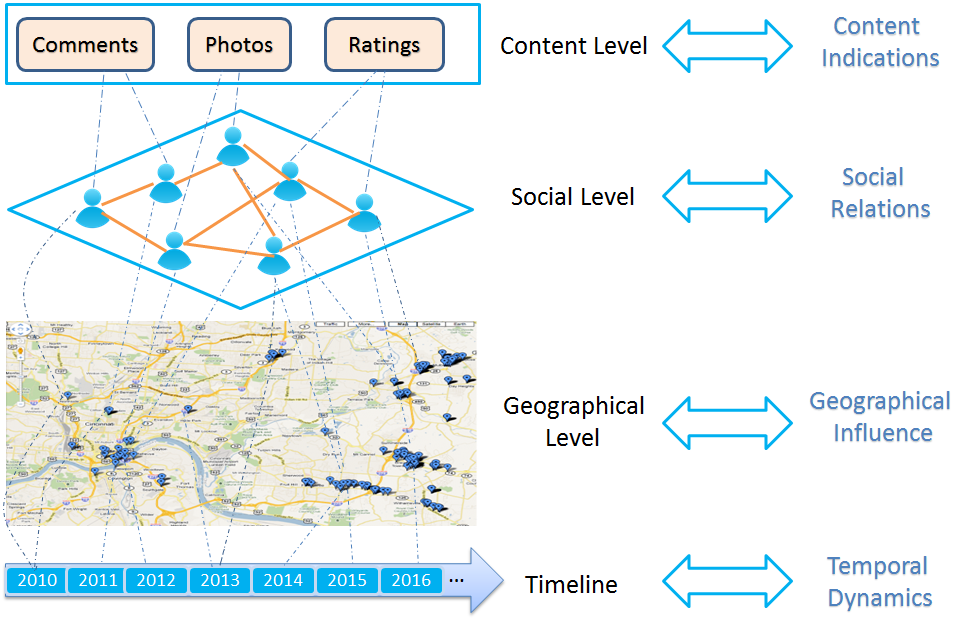}
\caption{Influential factors in LBSNs}
\label{fig:factor}
\end{figure}

We categorize the researches in POI recommendation according to several influential factors upon the user check-in activity. Because of the spatial and temporal properties resulted from the physical constraints and heterogeneous information such as locations' geographical information and users' comments, the check-in activity is a synthesized decision from a variety of factors. Figure~\ref{fig:factor} shows  four main factors in POI recommendations: geographical influence, temporal dynamics, social relations, and content indications. In the following, we demonstrate how each factor influences the check-in activity and how to model each influential factor for POI recommendation. 


\subsection{Geographical Influence}
\label{subsec:gi}
Geographical influence is an important factor
that distinguishes the POI recommendation
from traditional item recommendation, because
the check-in behavior depends on locations' geographical features. Analysis on users' check-in data show that, a user acts in geographically constrained areas and prefers to visiting POIs nearby those where the user has checked-in. 
Several studies~\cite{cheng2012fused,lian2014geomf,liu2014exploiting,ye2011exploiting,yuan2013time,zhang2013igslr,zhang2015geosoca,zhao2013capturing}  attempt to employ the geographical influence to improve POI recommendation systems. 
In particular, three representative models, i.e., power law distribution model, Gaussian distribution model, and kernel density estimation model, are proposed to capture the geographical influence in POI recommendation.

\begin{figure}[htbp]
\center
\includegraphics[scale=0.6]{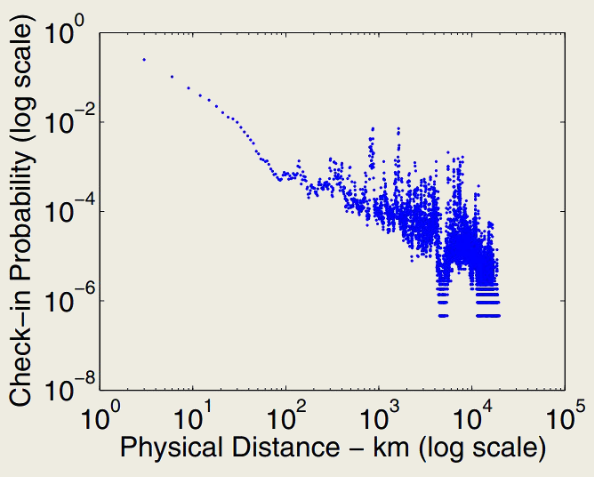}
\caption{Power law distribution pattern~\cite{ye2011exploiting}}
\label{fig:power}
\end{figure}

In~\cite{ye2011exploiting}, Ye et al. employ a power law distribution model to capture the geographical influence. Power law distribution pattern has been observed in human mobility such as withdraw activities in ATMs and travel in different cities~\cite{brockmann2006scaling,gonzalez2008understanding,rhee2011levy}. Also, Ye et al. discover similar pattern in users' check-in activity in LBSNs~\cite{ye2010location,ye2011exploiting}. Figure~\ref{fig:power} demonstrates two POIs' co-occurrence probability distribution over distance between two POIs. Because of the power law distribution in Figure~\ref{fig:power}, we are able to model the geographical influence as follows.
The co-occurrence 
probability $y$ of two POIs by the same user can be formulated as follows,
\begin{equation}
\label{eq:power}
y = a * x^b,
\end{equation}
where $x$ denotes the distance between two POIs, $a$ and $b$ are
parameters of the power-law distribution. Here, $a$ and $b$ should be learned from the observed check-in data, depicting the geographical feature of the check-in activity. A standard way to learn the parameters, $a$ and $b$, is to 
 transform Eq.~(\ref{eq:power}) to a linear equation via a logarithmic operation, and learn the parameters  by fitting a linear regression problem.

On basis of the geographical influence model depicted through the power law distribution, new POIs can be suggested according to the following formula. Given a past checked-in POI set $L_i$, the probability of visiting POI $l_j$ for user $u_i$, is formulated as,
\begin{equation}
Pr(l_j|L_i) = \frac{Pr(l_j \cup L_i)}{Pr(L_i)} = {\prod_{l_y \in L_i} Pr( d(l_j, l_y))},
\end{equation}  
where $d(l_j,l_y)$ denotes the distance between POI $l_j$ and $l_y$, and $Pr( d(l_j, l_y)) = a * d(l_j, l_y)^b$. In~\cite{ye2010location,ye2011exploiting}, Ye et al. leverage the power law distribution to model the geographical influence and combine it with  collaborative filtering techniques~\cite{ricci2011introduction} to recommend POIs. In addition, Yuan et al.~\cite{yuan2013time} also adopt the power law distribution model, but learn the parameter using a Bayesian rule instead. 

\begin{figure}[htbp]
\center
\includegraphics[scale=0.6]{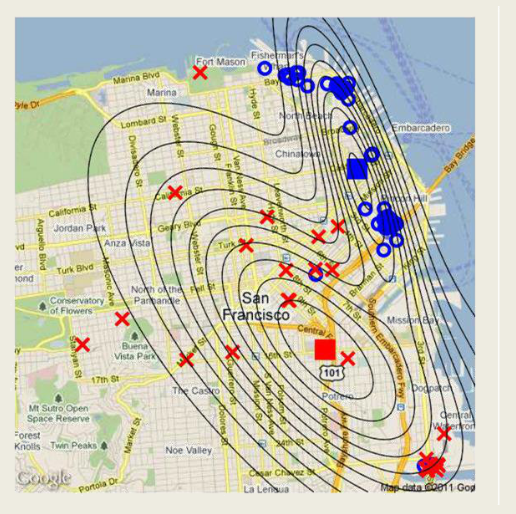}
\caption{Check-in distribution in multi-centers~\cite{cho2011friendship}}
\label{fig:mulcenter}
\end{figure}

The second type to model the geographical influence is a series of Gaussian distribution based methods.
Cho et al.~\cite{cho2011friendship} observe that users in LBSNs always act round some activity centers, e.g., home and office, as shown in Fig.~\ref{fig:mulcenter}. Further, Cheng et al.~\cite{cheng2012fused} propose a Multi-center Gaussian Model (MGM) to capture the geographical influence for POI recommendation. 
Given the multi-center set $C_u$, the probability of visiting POI $l$ by user $u$ is defined by
\begin{equation}
\label{eq:mgm}
P(l|C_u) = \sum_{c_u=1}^{|C_u|} P(l\in c_u) \frac{f_{c_u}^{\alpha}}{\sum_{i\in C_u} f_i^{\alpha}} \frac{N(l|{{\mu}_{C_u}, \sum_{C_u}})} {\sum_{i\in C_u} N(l|{\mu}_i, \sum_i)},
\end{equation}
where $P(l\in c_u)\propto \frac{1}{d(l,c_u)}$ is the probability of the POI $l$ belonging to the center $c_u$, $\frac{f_{c_u}^{\alpha}}{\sum_{i\in C_u} f_i^{\alpha}}$ denotes the normalized effect of the check-in frequency on the center $c_u$ and parameter $\alpha$ maintains the frequency aversion property, $N(l|{{\mu}_{C_u}, \sum_{C_u}})$ is the probability density function of Gaussian distribution with mean ${\mu}_{C_u}$ and covariance matrix $\sum_{C_u}$. Specifically, the MGM employs a greedy clustering algorithm on the check-in data to find the user activity centers. That may result in unbalanced assignment of POIs to different activity centers. Hence, Zhao et al.~\cite{zhao2013capturing} propose a genetic-based Gaussian mixture model to capture the geographical influence, which outperforms the MGM in POI recommendation.

\begin{figure}[htbp]
\center
\includegraphics[width=4.7in, height=1.6in]{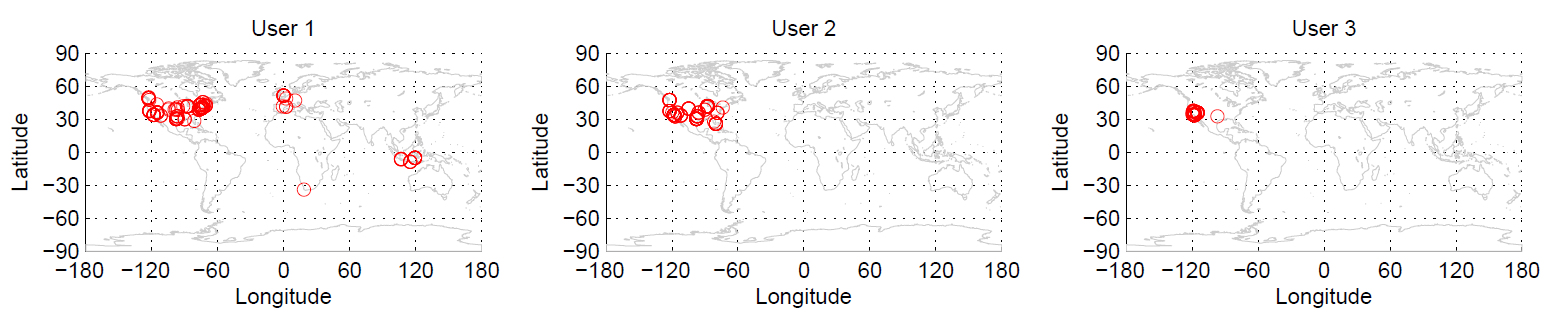}
\caption{Distributions of personal check-in locations~\cite{zhang2013igslr}}
\label{fig:kde}
\end{figure}

The third type of geographical model is the  kernel density estimation (KDE) model. 
In order to mine the personalized geographical influence, Zhang et al.~\cite{zhang2013igslr} argue that the geographical influence on each individual user should be  personalized rather than modeling though a common distribution, e.g., pow law distribution~\cite{ye2011exploiting} and MGM~\cite{cheng2012fused}. As shown in Fig.~\ref{fig:kde}, it is hard to model different users using the same distribution. To this end,  they leverage kernel density estimation~\cite{silverman1986density} to model the geographical influence using a personalized distance distribution for each user. Specifically, the kernel density estimation model consists of two steps: distance sample collection and distance distribution estimation. 
The step of distance sample collection generates a sample  $X_u$ for a user by computing the distance between every pair of locations visited by the user. 
Then, the distance distribution can be estimated through
the probability density function $f$  over distance $d$,
\begin{equation}
\label{eq:kde}
f(d) = \frac{1}{|X_u|\sigma} \sum_{d' \in X_u} K(\frac{d-d'}{\sigma}),
\end{equation}
where $\sigma$ is a smoothing parameter, called the bandwidth. $K(\cdot)$ is the Gaussian kernel
\begin{equation}
K(x) = \frac{1}{\sqrt{2\pi}}e^{-\frac{x^2}{2}}.
\end{equation}
Denote $L_u = \{l_1, l_2, \dots, l_n\}$ as the visited locations of user $u$. 
The probability of user $u$ visiting a new POI $l_j$ given the checked-in POI set $ L_u$ is defined as,
\begin{equation}
p(l_j|L_u) = \frac{1}{|L_u|} \sum_{l_i \in L_u} f(d_{ij}),
\end{equation}
where $d_{ij}$ is the distance between $l_i$ and $l_j$, $f(\cdot)$ is the distance distribution function in Eq.~(\ref{eq:kde}).
\subsection{Social Influence}
Inspired by the assumption that friends in LBSNs share more common interests than non-friends, social influence is explored to enhance POI recommendation~\cite{cheng2012fused,gao2012exploring,gao2012gscorr,ge2016point-of-interest,ye2010location,yang2013sentiment,zhang2015geosoca,zhang2014lore}. 
In fact, employing social influence to enhance recommendation systems has been explored in traditional recommendation systems, both in memory-based methods~\cite{jamali2009trustwalker,massa2007trust} and model-based methods~\cite{jamali2010matrix,ma2008sorec,ma2011recommender}. 
Researchers borrow the ideas from traditional recommendation systems to POI recommendation. In the following, we demonstrate representative researches capturing social influence in two aspects: memory-based and model-based. 

Ye et al.~\cite{ye2010location} propose a memory-based model, friend-based collaborative filtering (FCF) approach for POI recommendation. FCF model constrains the user-based collaborative filtering to find top similar users in friends rather than all users of LBSNs. Hence, the preference $r_{ij}$ of user $u_i$ at $l_j$ is calculated as follows,
\begin{equation}
{r}_{ij} = \frac{\sum_{u_k \in F_i} r_{kj}w_{ik}}{\sum_{u_k \in F_i}r_{kj}},
\end{equation}
where $F_i$ is the set of friends with top-$n$ similarity, $w_{ik}$ is similarity weight between $u_i$ and $u_k$. FCF enhances the efficiency by reducing the computation cost of finding top similar users. However, it overlooks the non-friends who share many common check-ins with the target user. Experimental results show that FCF brings very limited improvements over user-based POI recommendation in terms of precision. 

Cheng et al.~\cite{cheng2012fused} apply the probabilistic matrix factorization with social regularization (PMFSR)~\cite{ma2011recommender} in POI recommendation, which integrates social influence into PMF~\cite{salakhutdinov2008probabilistic}. Denote $\mathcal{U}$ and $\mathcal{L}$ are the set of users and POIs, respectively. PMFSR learns the latent features of users and POIs by minimizing the following objective function
\begin{equation}
\small
\argmin_{U,L} \sum_{i=1}^{|\mathcal{U}|} \sum_{j=1}^{|\mathcal{L}|} I_{ij} (g(c_{ij})-g(U_i^T L_j))^2 + \lambda_1 ||U||_F^2 + \lambda_2 ||V||_F^2 + \beta \sum_{i=1}^N \sum_{u_f \in F_i} sim(i,f) ||U_i-U_f||^2,
\end{equation}
where $U_i$, $U_f$, and $L_j$ are the latent features of user $u_i$, $u_f$, and POI $l_j$ respectively, $I_{ij}$ is an indicator denoting user $u_i$ has checked-in POI $l_j$, $F_i$ is the set of user $u_i$'s friends, $sim(i,f)$ denotes the social weight of user $u_i$ and $u_f$, and $g(\cdot)$ is the sigmoid function to mapping the check-in frequency value $c_{ij}$ into the range of [0,1]. In this framework, the social influence is incorporated by the social constraints that ensure latent features of friends keep in close distance at the latent subspace. Due to its validity, Yang et al.~\cite{yang2013sentiment} also employ the same framework to their sentiment-aware POI recommendation. 

\begin{figure}[htbp]
\center
\includegraphics[scale=0.5]{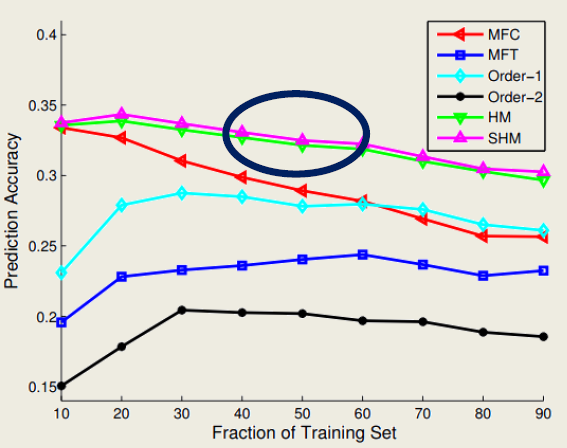}
\caption{The significance of social influence on POI recommendation~\cite{gao2012gscorr}}
\label{figsor}
\end{figure}
Although social influence improves traditional recommendation system significantly~\cite{jamali2010matrix,ma2008sorec,ma2011recommender}, the social influence on POI recommendation shows limited improvements~\cite{cheng2012fused,gao2012gscorr,ye2010location}. Figure~\ref{figsor} shows the limited improvement achieved from social influence in~\cite{gao2012gscorr}. Why this happens can be explained as follows.
Users in LBSNs make friends online without any limitation; on the contrary, the check-in activity requires physical interactions between users and POIs. Hence, friends in LBSNs may share common interest but may not visit common locations. For instance, friends in favour of Italian food from different cities will visit their own local Italian food restaurants. This phenomenon differs from the online movie and music recommendation scenarios such as Netflix and Spotify.

\subsection{Temporal Influence}
Temporal influence is of vital importance for POI recommendation because physical constraints on the check-in activity result in specific patterns.
Temporal influence  in a POI recommendation system performs in three aspects: periodicity, consecutiveness, and non-uniformness.

\begin{figure}[htbp]%
\centering
\subfigure[Day pattern]{
\includegraphics[width=.4\columnwidth, height = 1.8in]{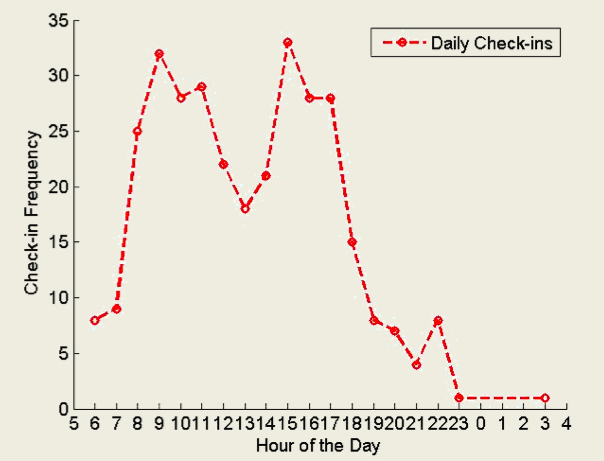}%
\label{subfig:day}%
}\hfill%
\subfigure[Week pattern]{
\includegraphics[width=.55\columnwidth , height = 1.8in]{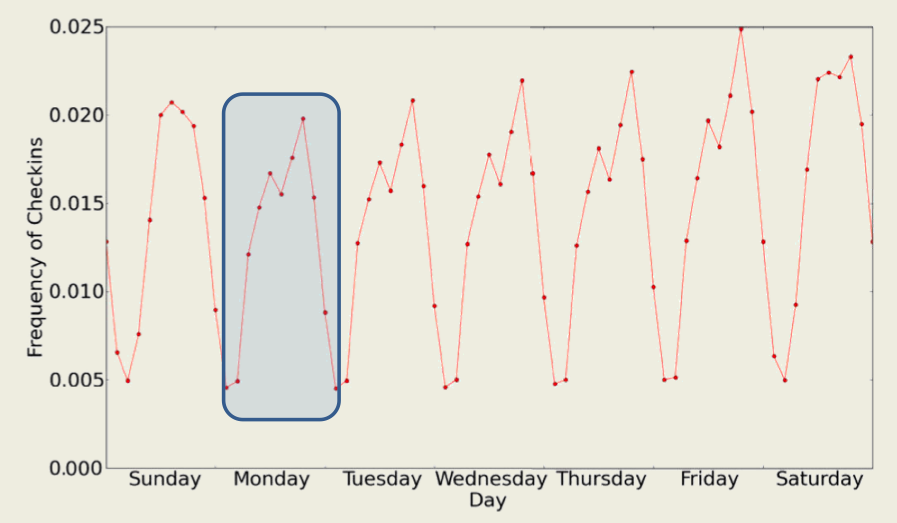}
\label{subfig:week}%
}
\caption{Periodic pattern~\cite{cheng2011exploring}}
\label{fig:period}
\end{figure}

Users' check-in behaviors in LBSNs exhibit periodic pattern. For instance, users always check-in restaurants at noon and have fun in nightclubs at night. Also users visit places around the office on weekdays and spend time in shopping malls on weekends. 
Figure~\ref{fig:period} shows the periodic pattern in a day and a week, respectively.  The check-in activity exhibits this kind periodic pattern, visiting the same or similar POIs at the same time slot. This observation inspires the researches exploiting this periodic pattern for POI recommendation~\cite{cho2011friendship,gao2013exploring,yuan2013time,zhang2015TICRec}.

\begin{figure}[htbp]%
\centering
\subfigure[CCDF of intervals in successive check-ins]{\includegraphics[height=1.6in, width=.45\columnwidth]{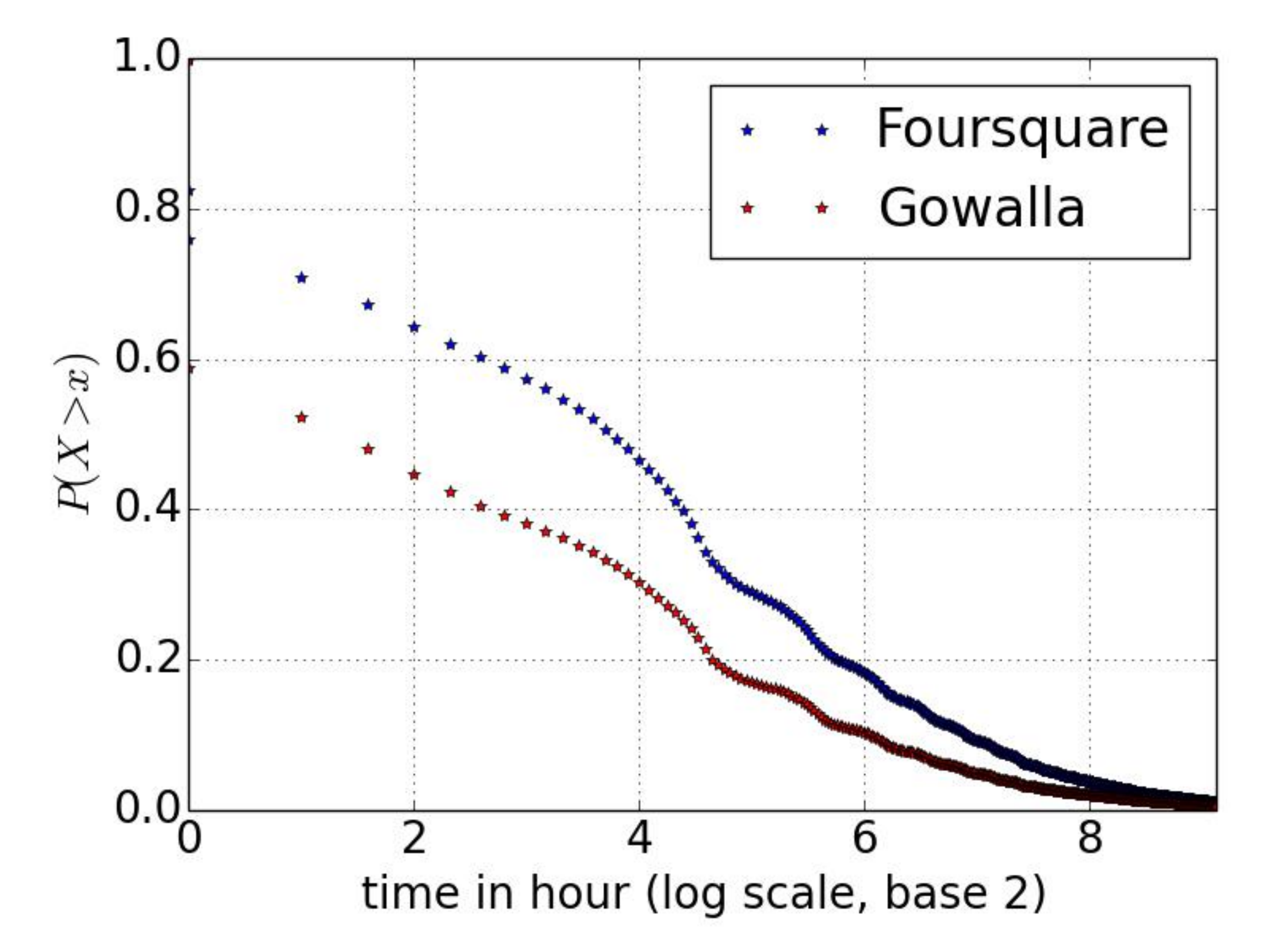}%
\label{subfig:intertime}%
}\hfill
\subfigure[CCDF of distances in successive check-ins]{
\includegraphics[height=1.6in,width=.45\columnwidth]{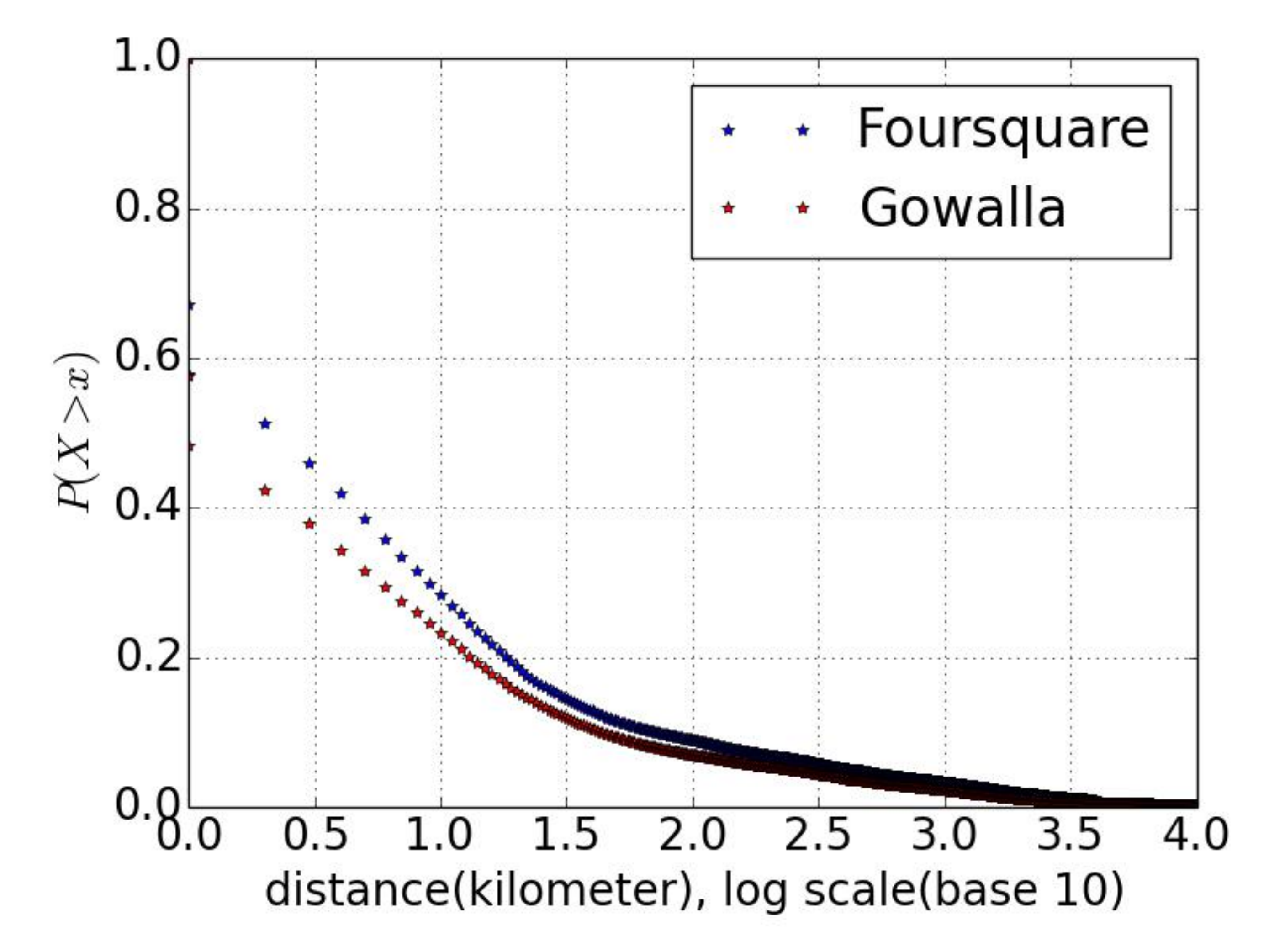}
\label{subfig:interpoi} }
\caption{Consecutive pattern~\cite{zhao2016stellar}}
\label{fig:inter_checkin}
\end{figure}

Consecutiveness performs in the check-in sequences, especially in the successive check-ins. 
Successive check-ins are usually correlated. For instance, users may have fun in a nightclub after diner in a restaurant. This frequent check-in pattern implies that the nightclub and the restaurant are geographically adjacent and correlated from the perspective of venue function. Data analysis on Foursquare and Gowalla in~\cite{zhao2016stellar} explores the spatial and temporal property of successive check-ins in Fig.~\ref{fig:inter_checkin}, namely, complementary cumulative distributive function (CCDF) of intervals and distances between successive check-ins. It is observed that many successive check-ins are highly correlated: over 40\% and 60\% successive check-in behaviors happen in less than 4 hours since last check-in in Foursquare and Gowalla respectively; about 90\% successive check-ins happen in less than 32 kilometers (half an hour driving distance) in Foursquare and Gowalla.
Researchers exploit Markov chain to model the sequential pattern~\cite{cheng2013you,feng2015personalized,he2016inferring,zhang2014lore}.
Researches in~\cite{cheng2013you,feng2015personalized} assume that two checked-in POIs in a short term are highly correlated and employ the factorized personalized Markov Chain (FPMC) model~\cite{rendle2010factorizing} to recommend successive POIs. Zhang et al.~\cite{zhang2014lore} propose an additive Markov model to learn the transitive probability between two successive check-ins. Zhao et al.~\cite{zhao2016stellar} exploit a latent factorization model to capture the consecutiveness, which is similar to the FPMC model in mathematical. 

\begin{figure}[htbp]
\center
\includegraphics[height=1.3in,width=.8\columnwidth]{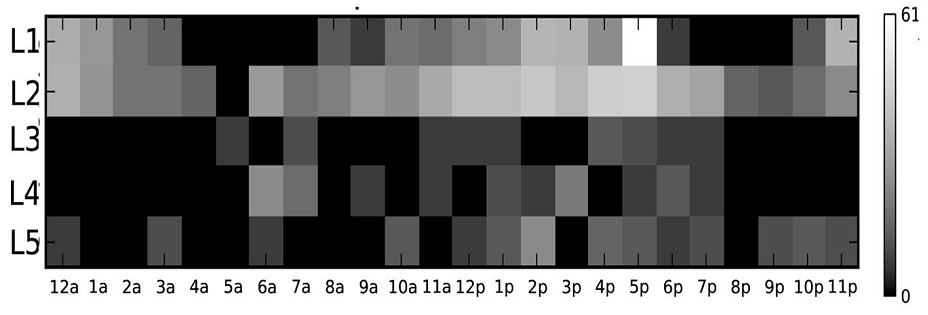}
\caption{Demonstration of non-uniformness~\cite{gao2012gscorr}}
\label{fig:nonuni}
\end{figure}

The non-uniformness feature depicts a user's check-in preference variance at different hours of a day, or at different months of a year, or at different days of a week~\cite{gao2013exploring}.  As shown in Fig.~\ref{fig:nonuni}, the study in~\cite{gao2013exploring} demonstrates an example of a random user's aggregated check-in activities on the user's top five most visited POIs. It is observed that a user's check-in preference changes at different hours of a day---the most frequent checked-in POI alters at different hours. 
 Similar temporal characteristics also appear at different months of a year, and different days of a week as well.
This non-uniformness feature can be explained from the user's daily life customs: 1) A user may check-in at POIs around the user's home in the morning hours, visit places around the office in the day hours, and have fun in bars in night hours. 2) A user may visit more locations around the user's home or office on weekdays. On weekends, the user may check-in more at shopping malls or vacation places. 3) At different months, a user may have different hobbies for food and entertainment. For instance, a user would visit ice cream shops in the months of summer while visit hot pot restaurants in the months of winter.

Although the temporal feature has been modeled to enhance the recommendation task, e.g., movie recommendation~\cite{koren2009collaborative,xiong2010temporal} and web service recommendation~\cite{zhang2011wspred}, the distinct temporal characteristics mentioned above make the previous temporal models unsatisfactory for POI recommendation. 
For example, the work in~\cite{koren2009collaborative} mines  temporal patterns of the Netflix data and incorporates the temporal influence into a matrix factorization model~\cite{koren2009matrix} to capture the user preference trends in a long range. The studies in~\cite{xiong2010temporal,zhang2011wspred} model the preference variance using a tensor factorization model. 
Since the previous proposed temporal models cannot meet the POI recommendation scenario, a variety of systems are proposed to enhance POI recommendation performance~\cite{cheng2013you,gao2013exploring,liu2016unified,yuan2013time,zhao2016stellar}

\subsection{Content Indication}
In LBSNs, users generate contents including tips and ratings for POIs and also photos about the POIs as well.
Although contents do not accompany each check-in record, the available contents, especially the user comments, can be used to enhance the POI recommendation~\cite{gao2015content,hu2014social,lian2015content,yang2013sentiment,yin2014lcars}. Because user comments provide extra information from the shared tips beyond the check-in behavior, e.g., the preference on a location. For instance, the check-in at an Italian restaurant does not necessarily mean the user likes this restaurant. Probably the user just likes Italian food but not this restaurant, even dislikes the taste of this restaurant. Compared with the check-in activity, the comments usually provide explicit preference information, which is a kind of complementary explanations for the check-in behavior. As a result, the comments are able to be used to deeply understand the users' check-in behavior and improve POI recommendation~\cite{gao2015content,hu2014social,yang2013sentiment}.    

\begin{figure}[htbp]
\center
\includegraphics[height=2.2in, width=.9\columnwidth]{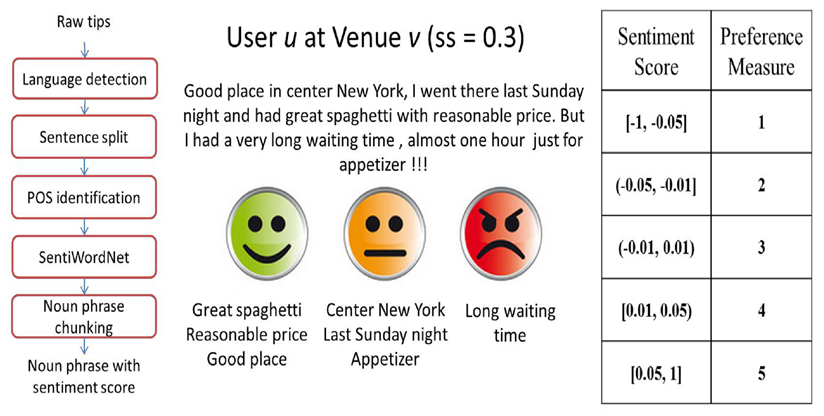}
\caption{Sentiment-preference transforming rule}
\label{fig:sent}
\end{figure}
The research in~\cite{yang2013sentiment} is the first and representative work exploiting the comments to strengthen the POI recommendation. Yang et al.~\cite{yang2013sentiment} propose a sentiment-enhanced location recommendation method, which utilizes the user comments to adjust the check-in preference estimation. 
As shown in Fig.~\ref{fig:sent}, the raw tips in LBSNs are collected and analysed using natural language processing techniques, including language detection, sentence split, POS identification, processed by SentiWordNet, and Noun phrase chunking. Then, each comment is given a sentiment score. According to the estimated sentiment, a preference score of one user at a POI is generated. 
Figure~\ref{fig:sent} also shows how to handle a comment example: transforming it to several noun phases such as ``Reasonable price", ``Good place", and ``Long waiting time", generating a sentiment score of 0.3, and mapping this value to the preference measure of 5. 
Moreover, through combining the preference measure from sentiment analysis and the check-in frequency, the proposed model in~\cite{yang2013sentiment} generates a modified rating $\hat{C}_{i,j}$ measuring the preference of user $u_i$ at a POI $l_j$. Accordingly, the traditional matrix factorization method can be employed to recommend POIs through the following objective,
\begin{equation}
\argmin_{U,L} \sum_{(i,j) \in \Omega} (\hat{C}_{i,j}-U_iL_j^T)^2 + \alpha ||U||_F^2 + \beta ||L||_F^2,
\end{equation} 
where $U_i$ and $L_j$ are latent features of user $u_i$ and $l_j$ respectively, $\hat{C}_{i,j}$ is the combined rating value, $\alpha$ and $\beta$ are regularizations.

\section{Taxonomy by Methodology}
\label{sec:tm}
In this section, we categorize the POI recommendation systems by the methodologies of using the influential factors mentioned above. 
In Sect.~\ref{sec:inf}, we discuss four general influential factors for POI recommendation. To establish a POI recommendation system requires to construct a model incorporating those influential factors. 

There are two ways to construct a POI recommendation system: the fused model and the joint model. The fused model  fuses recommended results from collaborative filtering method and recommended results from models capturing geographical influence, social influence, and temporal influence. The joint model establishes a joint model to learn the user preference and the influential factors together. 

\subsection{Fused Model}
The fused model usually establishes a model for each influential factor and combines their recommended results with suggestions from the collaborative filtering model~\cite{ricci2011introduction} that captures user preference on POIs. Since social influence provides limited improvements in POI recommendation and user comments are usually missing in users' check-ins, geographical influence and temporal influence constitute two important factors for POI recommendation. 
Hence, a typical fused model~\cite{cheng2012fused,ye2011exploiting,zhang2015geosoca} recommends POIs through combining the traditional collaborative filtering methods and influential factors, especially including geographical influence or temporal influence.

In~\cite{cheng2011exploring}, Cheng et al. employ probabilistic matrix factorization (PMF)~\cite{salakhutdinov2008probabilistic} and probabilistic factor model (PFM)~\cite{ma2011probabilistic} to learn user preference for recommending POIs. Suppose the number of users is $m$, and the number of POIs is $n$. $U_i$ and $L_j$ denote the latent feature of user $u_i$ and  POI $l_j$. PMF based method assumes Gaussian distribution on observed check-in data and Gaussian priors on the user latent feature matrix $U$ and POI latent feature matrix $L$. Then, the objective function to learn the model is as follows,
\begin{equation}
\label{eq:pmf}
\min_{U,L} \sum_{i=1}^{N} \sum_{j=1}^{M} I_{ij} (g(c_{ij})-g(U_i^TL_j))^2 + \lambda_1||U||_F^2 + \lambda_2 ||L||_F^2,
\end{equation} 
where $g(x)= \frac{1}{1+e^{-x}}$ is the logistic function, $c_{ij}$ is the checked-in frequency of user $u_i$ at POI $l_j$.  $I_{ij}$ is the indicator function to record the check-in state of $u_i$ at $l_j$. Namely, $I_{ij}$ equals one when the {$i$-th} user has checked-in at {$j$-th} POI; otherwise zero. 
After learning the user and POI latent features, the preference score of $u_i$ over $l_j$ is measured by the 
following score function,
\begin{equation}
\label{eq:score}
P(F_{ul}) = \sigma (U_i^TL_j),
\end{equation}
where $\sigma$ is the sigmoid function.

In addition, the geographical influence can be modeled through MGM, shown in Eq.~(\ref{eq:mgm}) of Sect.~\ref{subsec:gi}. 
Then, a fused model is proposed to combine user
preferences learned from Eq.~(\ref{eq:pmf}) and geographical influence modeled in Eq.~(\ref{eq:mgm}). The proposed model 
determines the probability $P_{ul}$ of a user $u$ visiting a location $l$  via the product of the preference socre estimation  
and the probability of whether a user
will visit that place in terms of geographical influence , 
\begin{equation}
P_{ul} = P(F_{ul}) \cdot P(l|C_u), 
\end{equation}
where $P(l|C_u)$ is calculated via the MGM and
$P(F_{ul})$ encodes a user’s preference on a location.
\subsection{Joint Model}
Different from the fused model, the joint model learns several influential factors together, and then recommends POIs from the jointly learned model. Compared with the fused model, a joint model connects different influential factors into the same final training target---the check-in behavior. The joint model depicts the check-in behavior as a synchronized decision influenced by several factors together, which better reflects the real scenario than the fused model.  This advantage over the fused model makes the joint model attract more attentions. Recently a number of joint models~\cite{gao2013exploring,gao2015content,hu2014social,kurashima2013geo,lian2014geomf,liu2013learning,liu2014exploiting,yang2013sentiment,yin2013lcars} are proposed for POI recommendation. The joint model contains two types: 1) incorporating factors (e.g., geographical influence and temporal influence) into traditional collaborative filtering model like matrix factorization and tensor factorization, e.g., \cite{gao2013exploring,gao2015content,lian2014geomf,liu2014exploiting,yang2013sentiment};  2) generating a graphical model according to the check-ins and extra influences like geographical information, e.g., \cite{hu2014social,liu2013learning,kurashima2013geo,yin2013lcars}.  
The key difference of the two types lies in different distribution assumptions on users' check-ins: the first type bases on collaborative filtering model that assumes Gaussian distribution while the second utilizes other types such as Possion distribution. 
\subsubsection{Representative Work for MF-based Joint Model}
In this section, we report two representative researches about the MF-based joint model, which incorporate temporal effect and geographical effect into a matrix factorization framework, respectively. 

In~\cite{gao2013exploring}, Gao et al. propose a Location Recommendation framework with Temporal effects (LRT), which incorporates temporal influence into a matrix factorization model. The LRT model contains two assumptions on temporal effect: 1) non-uniformness, users' check-in preferences change at different hours of one day; 2) consecutiveness, users' check-in preferences are similar in consecutive time slots. To model the non-uniformness, LRT separates a day into $T$ slots, and  defines time-dependent user latent feature $U_t \in R^{m \times d}$, where $m$ is the number of users, $d$ is the latent feature dimension, and $t \in [1, T]$ indexes time slots. Suppose that $C_t \in R^{m \times n}$ denotes a matrix depicting the check-in frequency at temporal state $t$. $U$ and $L$ denote the latent feature matrix for user and POI, respectively. Using the non-negative matrix factorization to model the POI recommendation system, the time-dependent objective function is as follows,
\begin{equation}
\label{eq:lrt}
\min_{U_t \geq 0, L \geq 0} \sum_{t=1}^T ||Y_t \odot (C_t - U_tL^T) ||_F^2 + \alpha \sum_{t=1}^T ||U_t||_F^2 + \beta ||L||_F^2,
\end{equation}   
where $Y_t$ is the corresponding indicator matrix, $\alpha$ and $\beta$ are the regularizations. Furthermore, the temporal consecutiveness inspires to minimize the following term,
\begin{equation}
\label{eq:lrtc}
\min \sum_{t=1}^{T}\sum_{i=1}^{m} \phi_i(t,t-1)||U_t(i,:)-U_{t-1}(i,:)||_2^2,
\end{equation}
where $\phi_i(t,t-1) \in [0,1]$ is defined as a temporal coefficient that measures user preference similarity between temporal state $t$ and $t-1$. The temporal coefficient could be calculated via cosine similarity according to users' check-ins at state $t$ and $t-1$. To represent the Eq.~(\ref{eq:lrtc}) in matrix form, we get
\begin{equation}
\min \sum_{t=1}^{T} Tr((U_t-U_{t-1})^T \Sigma_t (U_t-U_{t-1}),
\end{equation} 
where $\Sigma_t \in R^{m \times m}$ is the diagonal temporal coefficient matrix among $m$ users. Combining the two minimization targets, the objective function of the LRT model is gained as follows,
\begin{equation}
\begin{split}
\min_{U_t \geq 0, L \geq 0} \sum_{t=1}^T ||Y_t \odot (C_t - U_tL^T) ||_F^2 + \alpha \sum_{t=1}^T ||U_t||_F^2 + \beta ||L||_F^2\\
 +\lambda \sum_{t=1}^{T} Tr((U_t-U_{t-1})^T \Sigma_t (U_t-U_{t-1}),
\end{split}
\end{equation}
where $\lambda$ is a non-negative parameter to control the temporal regularization. User and location latent representations can be learned by solving the above optimization problem. Then, the user check-in preference $\hat{C}_t(i,j)$ at each temporal state can be estimated by the product of user latent feature and location feature ($U_t(i,:)L(j,:)^T$). Recommending POIs for users is to find POIs with higher value of  $\hat{C}(i,j).$ To aggregate different temporal states' contributions, $\hat{C}(i,j)$ is estimated through
\begin{equation}
\hat{C}(i,j) = f(\hat{C}_1(i,j), \hat{C}_2(i,j), \dots, \hat{C}_T(i,j)),
\end{equation}
where $f(\cdot)$ is an aggregation function, e.g., sum, mean, maximum, and voting operation.
 
\begin{figure}[htbp]
\center
\includegraphics[scale=0.8]{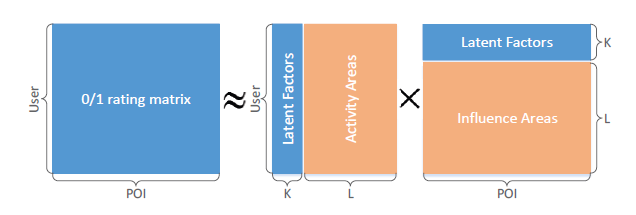}
\caption{Demonstration of GeoMF model~\cite{lian2014geomf}}
\label{fig:geomf}
\end{figure}

In~\cite{lian2014geomf}, Lian et al. propose the GeoMF model to incorporate geographical influence into a weighted regularized matrix factorization model (WRMF)~\cite{hu2008collaborative,pan2008one}.
WRMF is a popular model for one-class collaborative filtering problem, learning implicit feedback for recommendations. GeoMF treats the user check-in as implicit feedback and leverages a 0/1 rating matrix to represent the user check-ins. Furthermore, GeoMF employs an augmented matrix to recover the rating matrix, as shown in Fig.~\ref{fig:geomf}. Each entry in the rating matrix is the combination of two interactions: user feature and POI feature, users' activity area representation and POIs' influence area representation. Suppose there are $m$ users and $n$ POIs. The latent feature dimension is $d$ for user and POI representations, and is $l$ for users' activity area and POIs' influence area representations. Then the estimated rating matrix can be formulated as,
\begin{equation}
\label{eq:geomfscore}
\tilde{R} = PQ^T+ XY^T,
\end{equation}
where $\tilde{R} \in R^{m \times n}$ is the estimated matrix, $P\in R^{m \times d}$ and $Q \in R^{n \times d}$ are user latent matrix and POI latent matrix, respectively. In addition, $X \in R^{m \times l}$ and $Y \in R^{n \times l}$ are user activity area representation matrix and POI activity area representation matrix, respectively. Define $W$ as the binary weighted matrix whose entry $w_{ui}$ is set as follows, 
\begin{equation}
 w_{ui} =
  \begin{cases}
    \alpha(c_{ui}) + 1       & \quad \text{if } c_{ui} > 0 \\
    1  & \quad \text{otherwise, } \\
  \end{cases}
\end{equation}
where $c_{ui}$ is user $u$'s check-in frequency at POI $l_i$, $\alpha(c_{ui}) >0$ is a monotonically increasing function with respect to $c_{ui}.$
Following the scheme of WRMF model, the objective function of GeoMF is formulated as, 
\begin{equation}
\label{eq:geomf}
\argmin_{P,Q,X} ||W \odot (R-PQ^T-XY^T)||_F^2 + \gamma (||P||_F^2 + ||Q||_F^2) + \lambda ||X||_1,
\end{equation}
where $Y$ is  POIs' influence area matrix generated from a Gaussian kernel function, $P$, $Q$, and $X$ are parameters that need to learn, and $\gamma$ and $\lambda$ are regularizations.  After learning the latent features from Eq.~(\ref{eq:geomf}), the proposed model estimates the check-in possibility according to Eq.~(\ref{eq:geomfscore}), and then recommends the POIs with higher values for each user.

\subsubsection{Representative Work for Generative Graphical Model}
In this section, we report the representative research about the generative graphical model, which incorporates geographical influence into a generative graphical model.

\begin{figure}
\center
\includegraphics[scale=0.7]{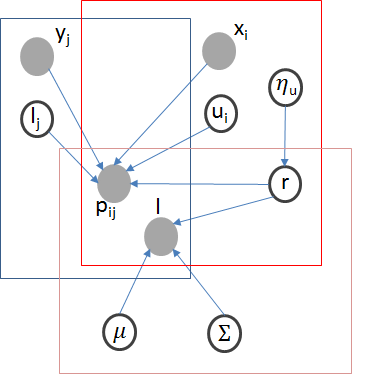}
\caption{A graphical representation of the model~\cite{liu2013learning} }
\label{fig:gmg}
\end{figure}

In~\cite{liu2013learning}, Liu et al. propose a geographical probabilistic factor analysis framework that takes various factors into consideration, including user preferences, the geographical influence, and the user mobility pattern. The proposed model mimics the user check-in decision process to
learn geographical user preferences for effective POI
recommendations. Figure~\ref{fig:gmg} demonstrates the graphical representation of the proposed model. Specifically, the proposed model assumes that the geographical locations have been clustered into several latent regions denoted as $R$. A multinomial distribution is applied to model user mobility over the regions $R$, $r \sim p(r|\eta_u)$, where $\eta_u$ is a user dependent distribution over latent regions for user $u_i$. Then, each region $r \in R$ is assumed to be a Gaussian geographical distribution and the POI $l_j$ is characterized by $l \sim \mathcal{N}(\mu_r, \sum_r)$ with $\mu_r$ and $\sum_r$ being the mean vector and covariance matrix of the region. In addition, the user check-in process is affected by the following factors: (1) each user $u_i$ is associated with an interest $\alpha(i,j)$ with respect to POI $l_j$; (2) each POI $l_j$ has popularity $\rho_j$; and (3) the distance between the user and the POI $d(u_i,l_j)$. Then, the probability of user $u_i$ visiting POI $l_j$ can be formulated as, 
\begin{equation}
p(u_i,l_j) \propto \alpha(i,j) \rho_j (d_0+d(u_i,l_j))^{-\tau},
\end{equation}
where a power-law like parametric term $(d_0+d(u_i,l_j))^{-\tau}$ is used to model the distance factor. Moreover, the user preference for POI can be represented as a linear combination of a latent factor $\textbf{u}_i^T\textbf{l}_j$ and a function of user and item observable properties $x^T_iWy_j$, namely 
\begin{equation}
\alpha(i,j) = \textbf{u}_i^T\textbf{l}_j + x^T_iWy_j.
\end{equation}
Because the proposed model uses implicit user check-in data to model user preferences and the distribution of check-in counts are usually skewed, a Bayesian probabilistic non-negative latent factor model is employed: 
$p_{ij} \sim P(f_{ij})$  where $f_{ij} = \alpha(i,j) \rho_j (d_0+d(u_i,l_j))^{-\tau}$.
Therefore, the proposed model shown in Fig.~\ref{fig:gmg} can be generated according to the following process:
\begin{enumerate}
 \item Draw a region $r \sim$ Multinomial($\eta_u$);
 \item Draw a location $l \sim \mathcal{N}(\mu_r, \sum_r)$;
 \item Draw a user preference
 \begin{enumerate}
 	\item Generate user latent factor $\textbf{u}_i \sim P(u_i;\Phi_{\textbf{u}})$;
 	\item Generate POI latent factor $\textbf{l}_j \sim P(\textbf{l}_j; \Phi_{\textbf{l}_j})$;
    \item User-item preference $\alpha(i,j) = \textbf{u}_i^T\textbf{l}_i + x^T_iWy_j$;
 \end{enumerate}
 \item $p_{ij} \sim P(f_{ij})$ where 
$ p_{ij} = (\textbf{u}_i^T\textbf{l}_j + x^T_iWy_j) \rho_j (d_0+d(u_i,l_j))^{-\tau}. $
\end{enumerate}

After the parameters are learned, the proposed model 
predicts the number of check-ins of a user for a given 
POI as $ \mathbb{E}(p_{ij}|u_i,l_j) = (\textbf{u}_i^T\textbf{l}_j + x^T_iWy_j) \rho_j (d_0+d(u_i,l_j))^{-\tau}. $ Moreover, POI recommendations are based on the predicted check-in times. The larger the predicted value is, the more likely the user will choose this POI.


\section{Taxonomy by Task}
\label{sec:tt}
In terms of whether to bias to the recent check-in, we categorize the POI recommendation task as general POI recommendation and successive POI recommendation. 
General POI recommendation in LBSNs is first proposed in~\cite{ye2010location}, which recommends the top-$N$ POIs for users, similar to movie recommendation task in Netflix competition. 
Further researches observe that two successive check-ins are significantly correlated in high probability, as shown in Fig.~\ref{fig:inter_checkin}.
Bao et al.~\cite{bao2012location} employ the recent check-in's information to recommend POIs for online scenario. Moreover, Cheng et al.~\cite{cheng2013you} propose the successive POI recommendation that provides favorite recommendations sensitive to the user's recent check-in. Namely, successive POI recommendation does not recommend users a general list of POIs but a list sensitive to their recent check-ins. Because successive POI recommendation takes advantage of the recent check-in information, it strikingly improves system performance on the recall metric. Hence, several studies~\cite{feng2015personalized,he2016inferring,zhang2015location,zhao2016stellar} are proposed for this specific POI recommendation task.

\subsection{General POI Recommendation}
The general POI recommendation task recommends the top-$N$ POIs for users, similar to movie recommendation task in Netflix competition. Researchers propose a variety of models to incorporate different influential factors, e.g., geographical influence and temporal influence, to fulfill this task~\cite{gao2013exploring,li2015rank,liu2013learning,ye2011exploiting}. In the following, we report a recent representative model for this task.

In~\cite{li2015rank}, Li et al. propose the geographical factorization method (Geo-FM), which employs the WARP loss to learn the recommended POI list. The check-in probability is assumed to be affected by two aspects: user preference and geographical influence, which are modeled by the interaction between the user and the target POI and the interaction between the user and neighboring POIs of the target POI. Further, a weight utility function is introduced to measure different neighbors' contribution in the geographical influence. For the neighbor $l'$ of target POI $l$, we set the weight $w_{l,l'} = (0.5 + d(l,l'))^{-1},$ where $d(l,l')$ denotes the distance between POI $l$ and $l'$. In practice, $w_{l,l'}$ may be normalized by devided by the sum of all values. Further, given user $u$ and POI $l$, we use $\textbf{u}_u^{(1)}$ and  $\textbf{u}_u^{(2)}$ to denote the user latent feature for user preference and geographical influence, and $\textbf{l}_l$ to denote the POI latent feature. 
Then, the recommendation score $y_{ul}$ could be formulated as,
\begin{equation}
\label{eq:geofme}
y_{ul} = \textbf{u}_u^{(1)} \cdot \textbf{l}_l + \textbf{u}_u^{(2)} \cdot  \sum_{l^* \in \mathcal{N}_k(l)} w_{l,l'} \cdot \textbf{l}_{l^*}, 
\end{equation}
where operator ($\cdot$) denotes the inner product, and $\mathcal{N}_k(l)$ denotes the $k$-nearest neighbors of POI $l$. 

After defining the recommendation score function, Geo-FM employs the WARP loss to learn the model. A user's preference ranking is summarized as that the higher the check-in frequency is, the more the POI is preferred by a user. In other words, for user $u$, POI $l$ would be ranked higher than $l'$ if $f_{ul} > f_{ul'}$, where $f_{ul}$ denotes the frequency of user $u$ at POI $l$. Given a user $u$ and a checked-in POI $l$, modeling the rank order is equivalent to minimize the following incompatibility, 
\begin{equation}
 Incomp(y_{ul}, \epsilon) = \sum_{ l,l' \in L,u \in U} I(f_{ul} > f_{ul'}) I(y_{ul} < y_{ul'} + \epsilon),
\end{equation}
where $U$ and $L$ denote the user set and POI set respectively, $\epsilon$ is the error tolerance hyperparameter,  and $I(\cdot)$ denotes the indicator function. 
 By modeling the incompatibility for all check-ins in the set $D$,  we get the objective function of the Geo-FM, 
\begin{equation}
\label{eq:geofm}
\mathcal{O} = \sum_{(u,l) \in D} E(Incomp(y_{ul}, \epsilon)),
\end{equation}
where $E(\cdot)$ is a function to convert the ranking incompatibility into a loss value: $E(r) = \sum_{i=1}^r \frac{1}{i}$.

Denote $L^C_u$ as the candidate POIs the user $u$ has not visited in POI set $L$. After learning the objective function in Eq.~(\ref{eq:geofm}), the check-in possibility of user $u$ over a candidate POI $l \in L^C_u$ could be estimated by Eq.~(\ref{eq:geofme}). Then, the POI recommendation task could be achieved through ranking the candidate POIs and selecting the top $N$ POIs with the highest estimated possibility values for each user.

\subsection{Successive POI Recommendation}
Successive POI recommendation, as a natural extension of general POI recommendation, is recently proposed and has attracted great research interest~\cite{cheng2012fused,feng2015personalized,zhang2015location,zhao2016stellar}. Different from general POI recommendation that focuses only on estimating users’ preferences on POIs, successive POI recommendation provides satisfied recommendations promptly based on users’ most recent checked-in location, which requires not only the preference modeling from users but also the accurate correlation analysis between POIs. In the following, we report a recent representative model for this task.  

In~\cite{zhao2016stellar}, Zhao et al. propose the STELLAR system, which aims to provide time-aware successive POI recommendations. The system attempts to rank the POIs via a score function $f: \mathcal{U} \times \mathcal{L} \times \mathcal{T}  \times \mathcal{L} \rightarrow \mathbb{R},$ which maps a four-tuple tensor to real values. Here, $\mathcal{U}$, $\mathcal{L}$, and $\mathcal{T}$ denote the set of users, the set of POIs, and the set of  time ids, respectively.
The score function $f(u, l^q, t, l^c)$ that represents the ``successive check-in possibility'', is defined for user $u$ to a candidate POI $l^c$ at the time stamp $t$ given the user's last check-in as a query POI $l^q$.

\begin{figure}[htbp]
\centering
\includegraphics[height=1.75in, width=5in]{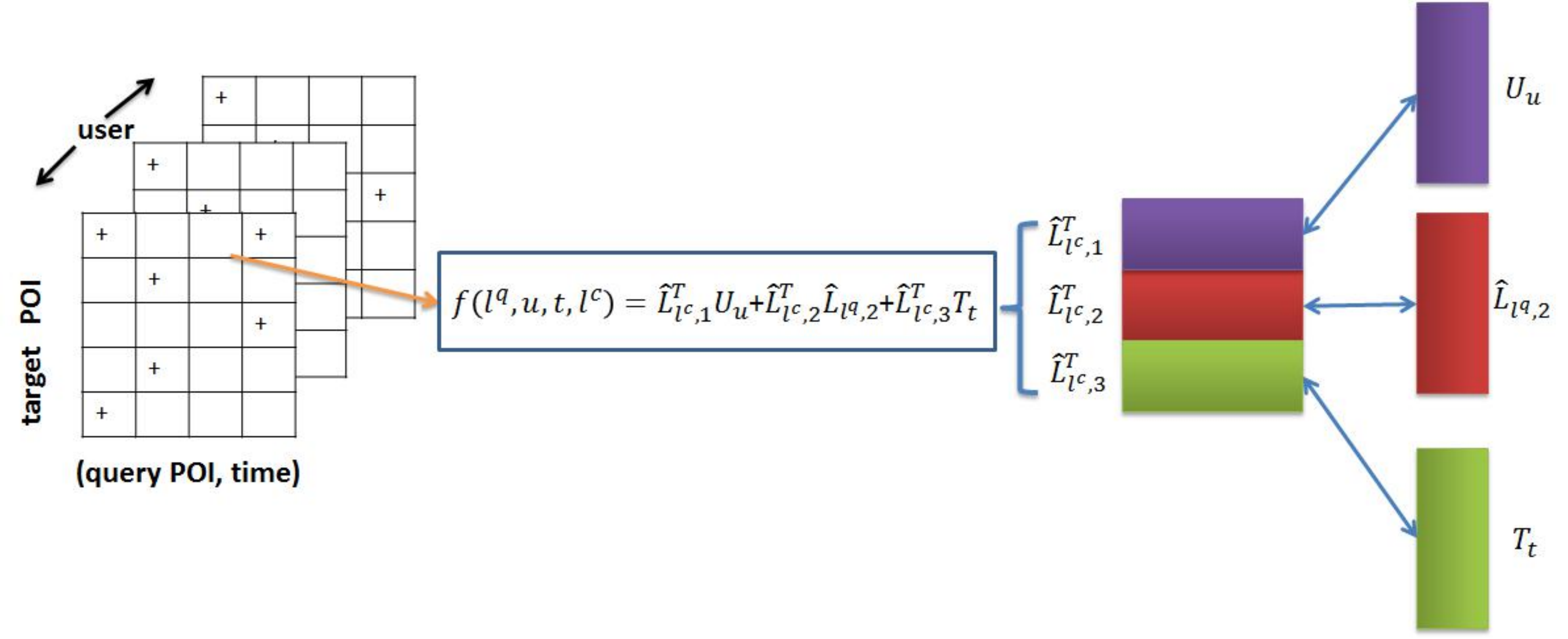}
\caption{STELLAR model formulation demonstration}
\label{fig:formulation}
\end{figure}

The STELLAR system learns the score function though a latent ranking framework.
Specifically, it employs pairwise tensor interactions to represent the following three key factors affecting users' check-in behavior: (1) the preference of user $u$ to a candidate POI $l^c$, (2) the temporal effect of  time  $t$ on a candidate POI $l^c$, and (3) correlation of the last checked-in POI $l^q$ and  a candidate POI $l^c$. Correspondingly, the score value of $f(u, l^q, t, l^c)$ is determined by user-POI interaction, time-POI interaction, and POI-POI interaction together.
Further, a $3 \times d$ matrix is used to represent POI latent feature, where $d$ is the latent space dimension. For each POI, three latent vectors are used to describe the POI-user, POI-time, and POI-POI interactions, respectively. As shown in Fig.~\ref{fig:formulation},  the function $f(u, l^q, t, l^c)$ is formulated as,
\begin{equation}
\label{eq:stellare}
f(u, l^q, t, l^c) = \hat{L}^T_{l^c,1}U_u  + \hat{L}^T_{l^c,2} \hat{L}_{l^q,2} + \hat{L}^T_{l^c,3}T_t.
\end{equation}
Here $U_u, T_t \in {R}^d$ are latent vectors of user {$u$} and time {$t$}, respectively; $\hat{L}_{l^c,1}$, $\hat{L}_{l^c,2}$, $\hat{L}_{l^c,3} \in {R}^d$ are candidate POI {$l^c$'s} three  {$d$-dimension} vectors  correspondingly interacting with users, other POIs, and time labels, respectively; and $ \hat{L}_{l^q,2}$ is query POI {$l^q$'s} latent vector interacting to the candidate POI. 
All latent vectors are set as {non-negative} to ensure better performance and real-world explanations on LBSNs for latent features. Then, the STELLAR system is made inference under BPR criteria. After obtaining the learning parameters, namely the latent feature matrices, the check-in possibility of user $u$ over a candidate POI $l \in L_u$ could be estimated by Eq.~(\ref{eq:stellare}). Then, the POI recommendation task could be achieved through ranking the candidate POIs and selecting the top $N$ POIs with the highest estimated possibility values for each user. Compared with the general POI recommendation task, successive POI recommendation is sensitive to the recent check-in. This is reflected in the check-in possibility estimation function: Eq.~(\ref{eq:stellare}) contains a query POI comparing with Eq.~(\ref{eq:geofme}).

\section{Performance Evaluation}
\label{sec:pe}

In this section, we report two important aspects for evaluating a POI recommendation system: data source and metrics. We first summarize several popular LBSN datasets. Then, we describe the metrics used to verify the effectiveness of the recommendation results. 
\subsection{Data Sources}
Gowalla, Brightkite, and Foursquare are famous benchmark datasets available for evaluating  a POI recommendation model. In this subsection, we briefly introduce these datasets and describe the statistics, shown in Table~\ref{tbl:data}. 

\begin{table}[h!]
\centering
\caption{LBSN datasets for POI recommendation}
\begin{tabular}
{p{3.5cm}  p{6cm}} \hline
Name  & Statistics \\ \hline
Brightkite~\cite{cho2011friendship} & 4,491,143 check-ins from 58,228 users \\ \hline
Gowalla~1~\cite{cho2011friendship} & 6,442,890 check-ins from 196,591 users \\ \hline
Gowalla~2~\cite{cheng2012fused} & 4,128,714 check-ins from 53,944 users \\ \hline
Foursquare~1~\cite{gao2012exploring} & 2,073,740 check-ins from 18,107 users \\ \hline
Foursquare~2~\cite{gao2012gscorr} & 1,385,223 check-ins from 11,326 users \\ \hline
Foursquare~3~\cite{bao2012location} & 325,606 check-ins from 80,606 users \\ \hline
\end{tabular}
\label{tbl:data}
\end{table}


\subsection{Metrics}
Most of POI recommendation systems utilize metrics of \textit{precision} and \textit{recall}, which are two general metrics to evaluate the model performance in information retrieval~\cite{davis2006relationship,goutte2005probabilistic}.  To see the balance of precision and recall,  \textit{F-score} is also introduced in some work. Since the precision and recall are low for POI recommendation, some researches~\cite{liu2013learning,ye2010location} introduce one relative metric, which measures the model comparative performance over random selection.

The precision and recall in the top-$N$ recommendation system are denoted as P@$N$ and R@$N$, respectively. P@$N$ measures the ratio of recovered POIs to the $N$ recommended POIs, and R@$N$ means the ratio of recovered POIs to the set of POIs in the testing data. For each user $u \in {U}$, ${L}^T_u$ denotes the set of correspondingly visited POIs in the test data, and  ${L}^R_u$ denotes the set of recommended POIs. Then,  the definitions of P@$N$ and R@$N$ are formulated as follows,
\begin{equation}
\label{eq:measure}
P@N = \frac{1}{|{U}|} \sum_{u \in {{U}}} \frac{|{L}^R_u \cap {L}^T_u|}{ N},
\end{equation}
\begin{equation}
\label{eq:measure}
R@N = \frac{1}{|{U}|} \sum_{u \in {{U}}} \frac{|{L}^R_u \cap {L}^T_u|}{|{L}^T_u|}.
\end{equation}
Further, \textit{F-score} is the harmonic mean of precision and recall. Therefore, the \textit{F-score} is defined as, 
\begin{equation}
\textit{F-score}@N = \frac{2*P@N * R@N}{P@N + R@N}.
\end{equation}   

In order to better compare the results, a relative metric is introduced. Relative precision@$N$ and recall@$N$ are denoted as r-P@$N$ and r-R@$N$, respectively.   Let $L^C_u$ denote the candidate POIs for each user $u$, namely POIs the user has not checked-in, then precision and recall of a random recommendation system is $\frac{|L^T_u|}{|L^C_u|}$ and $\frac{|N|}{|L^C_u|}$, respectively. Then, the relative precision@$N$ and recall@$N$ are defined as,
\begin{equation}
r-P@N = \frac{P@N}{{|L^T_u|}/{|L^C_u|}},
\end{equation}
\begin{equation}
r-R@N = \frac{R@N}{{|N|}/{|L^C_u|}}.
\end{equation}
\section{Trends and New Directions}
\label{sec:tnd}
In this section, we report the trends and new directions in POI recommendation. 
A bunch of studies have been proposed for POI recommendation. Summarizing the existing work, we point out the trends and new directions in two possible aspects: ranking-based model and online recommendation.
\subsection{Ranking-based Model}
Several ranking-based models~\cite{feng2015personalized,li2015rank,zhao2016stellar} have been proposed for POI recommendation recently. Most of previous methods generally attempt to estimate the user check-in probability over POIs~\cite{cheng2012fused,gao2013exploring,gao2015content}. However, for the POI recommendation task, we do not really care about the predicted check-in possibility value but the preference order. Some work has proved that it is better for the recommendation task to learn the order rather than the real value~\cite{rendle2009bpr,Lee2014Local,usunier2009ranking,weston2010large,weston2012latent1}. 
Bayesian personalized ranking (BPR) loss~\cite{rendle2009bpr} and weighted approximate rank pairwise  (WARP) loss~\cite{usunier2009ranking,weston2010large} are two popular criteria to learn the ranking order. Researchers in~\cite{feng2015personalized,zhao2016stellar} leverage the BPR loss to learn the model, and Li et al.~\cite{li2015rank} use the WARP loss. The existing work using ranking-base model has shown its advantage in model performance. Then, learning to rank, as an important technique for information retrieval~\cite{cao2007learning,liu2009learning}, may be used more for POI recommendation to improve performance in the future.

\subsection{Online Recommendation}
The online POI recommendation model has advantages over off-line models in two aspects: cold-start problem and adaptability to the user behavior variance. 
Most of previous work recommends POIs via the offline model, which suffers two problems: (1) cold-start problem, the offline model performs not satisfying for new users or users who have only a few check-ins; (2) user behaviour variance, the offline model may perform awfully if a user’s behaviour changes since it learns user behaviour according to history records.
Researchers in~\cite{bao2012location,yin2013lcars} utilize offline-model and online recommendation to improve the recommendation results. 
However, there is no work using online model for POI recommendation. 
In fact, online recommendation models based on multi-bandits have been proposed for movie recommendation and advertisement recommendation. In the future, online recommendation methods may be a new direction for POI recommendation. 

\section{Conclusion}
\label{sec:con}
Due to the prevalence of LBSNs and the importance of POI recommendation systems in LBNSs, we provide a systematic survey of the related recent researches. We review over 50 papers published in related top conferences and journals, including but not limited to AAAI, IJCAI, SIGIR, KDD, WWW, RecSys, UbiComp, ACM SIGSPATIAL, ACM TIST, and IEEE TKDE.
we categorize the systems by the influential factors, the methodology, and the task. Particularly we also report the representative work in each category.
This survey presents a panorama of this research with a balanced depth and scope. Further, this survey shows the trends and possible new directions in this area. 

\bibliographystyle{plain}
\bibliography{jref}  
\end{document}